\newfam\scrfam
\batchmode\font\tenscr=rsfs10 \errorstopmode
\ifx\tenscr\nullfont
        \message{rsfs script font not available. Replacing with calligraphic.}
        \def\scr{\cal}

\else   
        \font\sevenscr=rsfs7
        \font\fivescr=rsfs5
        \skewchar\tenscr='177 \skewchar\sevenscr='177 \skewchar\fivescr='177
        \textfont\scrfam=\tenscr \scriptfont\scrfam=\sevenscr
        \scriptscriptfont\scrfam=\fivescr
        \def\scr{\fam\scrfam}
        \def\cal{\scr}
\fi
\newfam\msbfam
\batchmode\font\twelvemsb=msbm10 scaled\magstep1 \errorstopmode
\ifx\twelvemsb\nullfont\def\Bbb{\bf}
        \message{Blackboard bold not available. Replacing with boldface.}
\else   \catcode`\@=11
        \font\tenmsb=msbm10 \font\sevenmsb=msbm7 \font\fivemsb=msbm5
        \textfont\msbfam=\tenmsb
        \scriptfont\msbfam=\sevenmsb \scriptscriptfont\msbfam=\fivemsb
        \def\Bbb{\relax\expandafter\Bbb@}
        \def\Bbb@#1{{\Bbb@@{#1}}}
        \def\Bbb@@#1{\fam\msbfam\relax#1}
        \catcode`\@=\active
\fi
        \font\eightrm=cmr8              \def\xrm{\eightrm}
        \font\eightbf=cmbx8             \def\xbf{\eightbf}
        \font\eightit=cmti10 at 8pt     \def\xit{\eightit}
        \font\eighttt=cmtt8             \def\xtt{\eighttt}
        \font\eightcp=cmcsc8
        \font\eighti=cmmi8              \def\xold{\eighti}
        \font\teni=cmmi10               \def\old{\teni}
        
        \font\tentt=cmtt10
        \font\twelverm=cmr12
        \font\twelvecp=cmcsc10 scaled\magstep1
        \font\fourteencp=cmcsc10 scaled\magstep2
        \font\fiverm=cmr5
        
        \font\eightmath=cmmi8

\def\noblackbox{\overfullrule=0pt}
\noblackbox

\def\SS{\scriptstyle}

\headline={\ifnum\pageno=1\hfill\else
{\eightcp T. Adawi, M. Cederwall, U. Gran, M. Holm, B.E.W. Nilsson:
        ``Superembeddings$\ldots$''}
                \dotfill{ }{\old\folio}\fi}
\def\makeheadline{\vbox to 0pt{\vss\noindent\the\headline\break
\hbox to\hsize{\hfill}}
        \vskip2\baselineskip}
\def\makefootline{
        \ifnum\foottest>0
                \ifnum\foottest=1
                        \footline={\footlineone}
                \fi
                \ifnum\foottest=2
                        \footline={\footlineone\footlinetwo}
                \fi
                \baselineskip=.8cm\vtop{\the\footline}
                \global\foottest=0
        \fi
        }
\newcount\foottest
\foottest=0
\def\footnote#1#2{${}^#1$\hskip-3pt
        \ifnum\foottest=1
        \def\footlinetwo{\hfill\break
        \vtop{\baselineskip=9pt
        \indent ${}^#1$ \vtop{\hsize=14cm\noindent\xrm #2}}}\foottest=2
        \fi
        \ifnum\foottest=0
        \def\footlineone{\vtop{\baselineskip=9pt
        \hrule width.6\hsize\hfill\break
        \indent ${}^#1$ \vtop{\hsize=14cm\noindent\xrm #2}}
        \vskip-.7\baselineskip}
        \foottest=1
        \fi
        }
\newcount\refcount
\refcount=0
\newwrite\refwrite
\def\ref#1#2{\global\advance\refcount by 1
        \xdef#1{{\old\the\refcount}}
        \ifnum\the\refcount=1
        \immediate\openout\refwrite=\jobname.refs
        \fi
        \immediate\write\refwrite
                {\item{[{\xold\the\refcount}]} #2\hfill\par\vskip-2pt}}
\def\refout{\catcode`\@=11
        \xrm\immediate\closeout\refwrite
        \vskip2\baselineskip
        {\noindent\twelvecp References}\hfill\vskip\baselineskip
        \baselineskip=.75\baselineskip
        \input\jobname.refs
        \baselineskip=4\baselineskip \divide\baselineskip by 3
        \catcode`\@=\active\rm}
\newcount\sectioncount
\sectioncount=0
\def\section#1#2{\global\eqcount=0
        \global\advance\sectioncount by 1
        \vskip2\baselineskip\noindent
        \hbox{\twelvecp\the\sectioncount. #2\hfill}\vskip\baselineskip\noindent
        \xdef#1{\the\sectioncount}}
\newcount\appendixcount
\appendixcount=0
\def\appendix#1{\global\eqcount=0
        \global\advance\appendixcount by 1
        \vskip2\baselineskip\noindent
        \ifnum\the\appendixcount=1
        \hbox{\twelvecp Appendix A: #1\hfill}\vskip\baselineskip\noindent\fi
    \ifnum\the\appendixcount=2
        \hbox{\twelvecp Appendix B: #1\hfill}\vskip\baselineskip\noindent\fi
    \ifnum\the\appendixcount=3
        \hbox{\twelvecp Appendix C: #1\hfill}\vskip\baselineskip\noindent\fi}
\def\acknowledgements{\vskip2\baselineskip\noindent
        \underbar{\it Acknowledgements:}\ }
\newcount\eqcount
\eqcount=0
\def\Eqn#1{\global\advance\eqcount by 1
        \xdef#1{{\old\the\sectioncount}.{\old\the\eqcount}}
        \ifnum\the\appendixcount=0
                \eqno({\oldstyle\the\sectioncount}.{\oldstyle\the\eqcount})\fi
        \ifnum\the\appendixcount=1
                \eqno({\oldstyle A}.{\oldstyle\the\eqcount})\fi
        \ifnum\the\appendixcount=2
                \eqno({\oldstyle B}.{\oldstyle\the\eqcount})\fi
        \ifnum\the\appendixcount=3
                \eqno({\oldstyle C}.{\oldstyle\the\eqcount})\fi}
\def\eqn{\global\advance\eqcount by 1
        \ifnum\the\appendixcount=0
                \eqno({\oldstyle\the\sectioncount}.{\oldstyle\the\eqcount})\fi
        \ifnum\the\appendixcount=1
                \eqno({\oldstyle A}.{\oldstyle\the\eqcount})\fi
        \ifnum\the\appendixcount=2
                \eqno({\oldstyle B}.{\oldstyle\the\eqcount})\fi
        \ifnum\the\appendixcount=3
                \eqno({\oldstyle C}.{\oldstyle\the\eqcount})\fi}
\def\multi{\global\advance\eqcount by 1}
\def\multieq#1#2{\xdef#1{{\old\the\eqcount#2}}
        \eqno{({\oldstyle\the\eqcount#2})}}
\parskip=3.5pt plus .3pt minus .3pt
\baselineskip=14pt plus .1pt minus .05pt
\lineskip=.5pt plus .05pt minus .05pt
\lineskiplimit=.5pt
\abovedisplayskip=18pt plus 4pt minus 2pt
\belowdisplayskip=\abovedisplayskip
\hsize=15cm
\vsize=19cm
\hoffset=1cm
\voffset=1.8cm
\def\ss{\scriptscriptstyle}
\def\hr{\hfill\cr}
\def\\{\cr}
\def\/{\over}
\def\*{\partial}
\def\a{\alpha}
\def\b{\beta}
\def\c{\gamma}
\def\d{\delta}
\def\e{\varepsilon}

\def\g{\gamma}

\def\m{\mu}

\def\th{\theta}
\def\w{\omega}
\def\W{\Omega}
\def\ki{\chi}

\def\G{\Gamma}

\def\punkt{\,\,.}
\def\komma{\,\,,}
\def\minus{\!-\!}
\def\+{\!+\!}
\def\={\!=\!}
\def\half{{\lower2.5pt\hbox{\eightrm 1}\/\raise2.5pt\hbox{\eightrm 2}}}
\def\fraction#1{{\lower2.5pt\hbox{\eightrm 1}\/\raise2.5pt\hbox{\eightrm #1}}}
\def\Fraction#1#2{
        {\lower2.5pt\hbox{\eightrm #1}\/\raise2.5pt\hbox{\eightrm #2}}}
\def\ihalf{{\lower2.5pt\hbox{\eightmath i}\/\raise2.7pt\hbox{\eightrm 2}}}
\def\ifrac#1{{\lower2.5pt\hbox{\eightmath i}\/\raise2.5pt\hbox{\eightrm#1}}}
\def\tr{\hbox{\rm tr}}

\def\epil{\hookrightarrow}

\def\stack#1#2{\buildrel {#1}\over {#2}}

\def\ul{\underline}
\def\ol{\overline}
\def\bl{\bar}

\def\ws{{\cal M}}
\def\wm{h}
\def\we{e}
\def\ww{\w}
\def\wT{{\cal T}}
\def\wr{{\cal R}}
\def\wd{d}
\def\wco{\th}
\def\wdel{{\cal D}}

\def\wtco{\tilde{\th}}
\def\cl{{\cal L}}

\def\im{g}
\def\iE{E}
\def\iw{\W}
\def\iT{T}
\def\ir{R}
\def\id{d}
\def\ico{\th}
\def\idel{\nabla}

\def\ec{K}
\def\cec{{\cal K}}
\def\eT{T}

\def\nm{g'}
\def\nE{E}
\def\nw{\W}
\def\nT{T}
\def\nr{R}
\def\nd{d'}
\def\nco{\th'}
\def\ndel{\nabla'}

\def\ts{\ul{\cal M}}
\def\tm{\ul{g}}
\def\tE{E}
\def\tw{\W}
\def\tT{T}
\def\tmT{\ul{T}}
\def\tr{R}

\def\td{\ul{d}}
\def\tco{\ul{\th}}
\def\tdel{\ul{\nabla}}
\def\te{{\cal E}}

\def\um{\ul{m}}
\def\un{\ul{n}}

\def\ioldel{\ol{\nabla}}
\def\woldel{\ol{\cal D}}
\def\woldelmod{\hat{\ol{\cal D}}}

\def\inm{(\!m^{\ss\minus 1}\!)}
\def\inte{(\!\te^{\ss\minus 1}\!)}

\def\ta{{\ul{a}}}
\def\tb{{\ul{b}}}
\def\tc{{\ul{c}}}
\def\td{{\ul{d}}}

\def\ba{{\ol{a}}}
\def\bb{{\ol{b}}}

\def\taa{{\ul{\a}}}
\def\tbb{{\ul{\b}}}
\def\tcc{{\ul{\c}}}

\def\baa{{\ol{\a}}}
\def\bbb{{\ol{\b}}}

\def\tA{{\ul{A}}}
\def\tB{{\ul{B}}}
\def\tC{{\ul{C}}}

\def\bA{{\ol{A}}}
\def\bB{{\ol{B}}}
\def\bC{{\ol{C}}}

\def\ttm{{\ul{m}}}

\def\tmm{{\ul{\mu}}}

\def\tM{{\ul{M}}}

\def\Gauss{Gauss--Weingarten}

\def\mx#1{\left(\matrix{#1}\right)}
\def\tbl#1{\matrix{#1}}
\def\eq#1{$$#1\eqn$$}
\def\eql#1#2{$$#1\Eqn#2$$}
\def\eqa#1{$$\eqalign{#1}\eqn$$}

%
%
%

\null\vskip-1cm
\hbox to\hsize{\hfill G\"oteborg-ITP-97-15}
\hbox to\hsize{\hfill\tt hep-th/9711203}
\hbox to\hsize{\hfill November, 1997}

\vskip3cm
\centerline{\fourteencp Superembeddings,
        Non-linear Supersymmetry and 5-branes}
\vskip4pt
\vskip\parskip
\centerline{\twelvecp}

\vskip1.2cm
\centerline{\twelverm Tom Adawi, Martin Cederwall, Ulf Gran,}
\centerline{\twelverm Magnus Holm and Bengt E.W. Nilsson}

\vskip.8cm
\centerline{\it Institute for Theoretical Physics}
\centerline{\it G\"oteborg University and Chalmers University of Technology }
\centerline{\it S-412 96 G\"oteborg, Sweden}

\vskip.8cm
\catcode`\@=11
\centerline{\tentt adawi,tfemc,gran,holm,tfebn@fy.chalmers.se}
\catcode`\@=\active

\vskip2.2cm

\centerline{\bf Abstract}

{\narrower\noindent We examine general properties of superembeddings,
i.e., embeddings of supermanifolds into supermanifolds. The connection between
an embedding procedure and the method of non-linearly realised supersymmetry
is clarified, and we demonstrate how the latter arises as a special
case of the former. As an illustration, the super-5-brane in 7 dimensions,
containing a self-dual 3-form world-volume field strength,
is formulated in both languages, and provides an example of a model
where the embedding condition does not suffice to put the theory on-shell.
\smallskip}

\vfill
\eject

\def\nl{\hfill\break\indent}
\def\nlni{\hfill\break}

\ref\Review{C.M. Hull and P.K. Townsend,
        {\xit ``Unity of superstring dualities''},
        Nucl.Phys. {\xbf B438} ({\xold1995}) {\xold109}
        [{\xtt hep-th/9410167}];\nlni
        E. Witten, {\xit ``String theory dynamics in various dimensions''},
        Nucl.Phys. {\xbf B443} ({\xold1995}) {\xold85}
        [{\xtt hep-th/9503124}];\nlni
        J.H. Schwarz, {\xit ``The power of M theory''},
        Phys.Lett. {\xbf B367} ({\xold1996}) {\xold97}
        [{\xtt hep-th/9510086}];\nlni
        P.K. Townsend, {\xit ``Four lectures on M-theory''},
        {\xtt hep-th/9612121}.}
\ref\CTW{P.K. Townsend, {\xit ``Membrane tension and manifest IIB S-duality''},
        {\xtt hep-th/9705160};\nlni
        M. Cederwall and P.K. Townsend,
        {\xit ``The manifestly Sl(2;Z)-covariant superstring''},\nl
        JHEP {\xbf09} ({\xold1997}) {\xold003} [{\xtt hep-th/9709002}];\nlni
        M. Cederwall and A. Westerberg,
        {\xit ``World-volume fields, SL(2;Z) and duality:
        the type IIB 3-brane''}, \nl{\xtt hep-th/9710007}.}
\ref\Branes{M.J. Duff, {\xit ``Supermembranes''}, {\xtt hep-th/9611203};\nlni
        J. Polchinski, {\xit ``TASI lectures on D-branes''},
        {\xtt hep-th/9611050}.}
\ref\HSWFivebrane{P.S.~Howe and E.~Sezgin, {\xit ``D=11, p=5''},
        Phys.~Lett.~{\xbf B394} ({\xold1997}) {\xold62}
        [{\xtt hep-th/9611008}];\nlni
        P.S.~Howe, E.~Sezgin and P.C.~West,\nl
        {\xit ``Covariant field equations of the M theory five-brane''},
        Phys.~Lett.~{\xbf B399} ({\xold1997}) {\xold49}
        [{\xtt hep-th/9702008}];\nl
        {\xit ``The six-dimensional self-dual tensor''},
        Phys.~Lett.~{\xbf B400} ({\xold1997}) {\xold255}
        [{\xtt hep-th/9702111}]}
\ref\Fivebrane{I. Bandos, K. Lechner, A. Nurmagambetov, P. Pasti, D. Sorokin
        and M. Tonin,\nl {\xit ``Covariant action for the super-five-brane
                of M-theory''},\nl
        Phys. Rev. Lett. {\xbf78} ({\xold1997}) {\xold4332}
                [{\xtt hep-th/9701037}];\nlni
        M. Aganagic, J. Park, C. Popescu and J.H. Schwarz,
        {\xit ``World-volume action of the M theory five-brane''},
        \nl Nucl. Phys. {\xbf B496} ({\xold1997}) {\xold191}
                [{\xtt hep-th/9701166}].}
\ref\Bagger{J.~Bagger and A.~Galperin,
        {\xit ''Matter couplings in partially broken extended supersymmetry''},
        \nl Phys. Lett. {\xbf B336} ({\xold1994}) {\xold25}
        [{\xtt hep-th/9406217}];\nl
        {\xit ``A new Goldstone multiplet for partially broken
        supersymmetry''},\nl
        Phys.~Rev.~{\xbf D55} ({\xold1997}) {\xold1091}
        [{\xtt hep-th/9608177}];\nl
        {\xit ``The tensor Goldstone multiplet for partially broken
        supersymmetry''}, {\xtt hep-th/9707061}.}
\ref\PST{I.A.~Bandos, P.~Pasti, D.~Sorokin, M.~Tonin and D.V.~Volkov,
        {\xit ``Superstrings and supermembranes in the doubly\nl supersymmetric
        geometrical approach''},
        Nucl.~Phys.~{\xbf B446} ({\xold1995}) {\xold79}
        [{\xtt hep-th/9501113}];\nlni
        I.A. Bandos, D. Sorokin and D.V. Volkov, \nl{\xit ``On the generalized
        action principle for superstrings and supermembranes''},
        Phys. Lett. {\xbf B352} ({\xold1995}) {\xold269};\nlni
        I.A.~Bandos,
        {\xit ``On a zero curvature representation for bosonic strings and
        p-branes''},\nl
        Phys.Lett.~{\xbf B388} ({\xold1996}) {\xold35}
        [{\xtt hep-th/9510216}];\nlni
        I.A.~Bandos and W.~Kummer,
        {\xit ``p-branes, Poisson-sigma-models and embedding approach to
        (p+1)-dimensional\nl gravity''},
        {\xtt hep-th/9703099};\nlni
        I.A.~Bandos, P.~Pasti, D.~Sorokin and M.~Tonin,\nl
        {\xit ``Superbrane actions and geometrical approach''},
        {\xtt hep-th/9705064};\nlni
        I.A.~Bandos and W.~Kummer,
        {\xit ``A polynomial first order action for the dirichlet 3-brane''},
        {\xtt hep-th/9707110}.}
\ref\Sezgin{E. Bergshoeff and E.~Sezgin,
        {\xit ``Twistor-like formulation of super p-branes''},
        \nl Nucl.~Phys.~{\xbf B422} ({\xold1994}) {\xold329}
	[{\xtt hep-th/9312168}];\nlni
        E.~Sezgin,
        {\xit ``Spacetime and worldvolume supersymmetric super p-brane
        actions''},
        {\xtt hep-th/9411055}.}
\ref\HSW{
        P.S.~Howe and E.~Sezgin,
        {\xit ``Superbranes''},
        Phys.~Lett.~{\xbf B390} ({\xold1997}) {\xold133}
        [{\xtt hep-th/9607227}];\nlni
        P.S.~Howe, E.~Sezgin and P.C.~West,
        {\xit ``Aspects of superembeddings''},
        {\xtt hep-th/9705093}.}
\ref\Kallosh{R. Kallosh, {\xit ``Volkov-Akulov theory and D-branes''},
        {\xtt hep-th/9705118}.}
\ref\Kobayashi{S. Kobayashi and K. Nomizu, {\xit ``Foundations of differential
        geometry, vol. II''} (Wiley Interscience, 1963).}
\ref\Volkov{V.I. Ogievetskii, in Xth Karpacz winter school of theoretical
        physics (1974).}
\ref\Akulov{D.V. Volkov and V.P. Akulov, JETP Lett. {\xbf16} ({\xold1972})
        {\xold438}.}
\ref\Strathdee{J. Strathdee, {\xit ``Extended Poincar\'e supersymmetry''},
        Int. J. Mod. Phys. {\xbf A2} ({\xold1987}) {\xold273}.}
\ref\HST{P.S.~Howe, G.~Sierra and P.K.~Townsend,
                {\xit ``Supersymmetry in six dimensions''},
                Nucl.~Phys.~{\xbf B221} ({\xold1983}) {\xold331}.}
\ref\SDLagrangian{P. Pasti, D. Sorokin and M. Tonin,
        {\xit''On Lorentz invariant actions for chiral p-forms''},\nl
        Phys.Rev. {\xbf D55} ({\xold1997}) {\xold6292}
        [{\xtt hep-th/9611100}].}
\ref\Gates{S.J.~Gates, Jr. and W.~Siegel,
        {\xit ``Understanding constraints in superspace formulations of
        supergravity''},\nl
        Nucl.~Phys.~{\xbf B163} ({\xold1980}) {\xold519}.}
\ref\KT{T.~Kugo and P.~Townsend,
        {\xit ``Supersymmetry and the division algebras''},
        Nucl.~Phys.~{\xbf B221} ({\xold1983}) {\xold357}.}

\frenchspacing


\section\Intro{Introduction}Our understanding of string theory
at the non-perturbative level has gone through a
dramatic improvement in recent years. Some of the key aspects
of this development are connected to the central r\^ole played by
solitonic solutions of the low energy field equations,
i.e., various brane configurations that solve the field equations
of the supergravity theories. By considering BPS saturated solitonic
solutions which preserve e.g. half the supersymmetry of the supergravity
theory in question, these supergravity theories can be shown to be related by
duality transformations some of which are intrinsically non-perturbative
in nature. In fact, (almost)
all consistent string and supergravity theories, including 11-dimensional
supergravity, are in this way believed
to constitute low-energy descriptions of one master theory, the so called
M-theory, in either the weak or strong coupling
regime of some particular coupling constant in the moduli space of
all couplings. An overview of the subject, as well as further references,
may be found e.g. in ref. [\Review].

The known branes come in three main varieties\footnote\dagger{There are also
branes associated with gravitational charges. We will not consider these
in the present paper.},
$p$-branes, D$p$-branes, and T5-branes, depending on whether the bosonic sector
of the field theory
on the world-volume of the brane contains only scalars, scalars and
vector gauge fields, or scalars together with a third rank anti-symmetric
self-dual tensor field strength. (Recently also other types of
tensor fields and combination of
such have been introduced in these theories to solve certain
specific problems [\CTW]. However, this is
of no immediate interest for the considerations of this paper).
For a review of the different kinds of solitonic branes and their
r\^oles in non-perturbative string theory, see ref. [\Branes].
The scalar fields appearing on the branes are immediately identifyable
as Goldstone fields, or collective modes, corresponding to the
translation symmetries that are broken when the brane is introduced
into target space-time. That is, one obtains one
scalar field for each direction transverse to
the brane. By checking which supersymmetries get broken, or by viewing
the brane as a supersurface embedded in a target superspace, also the number
of Goldstone fermions can be deduced.
However, when supersymmetry requires
the brane supermultiplet to contain also vectors or tensor potentials, there
is no analogously simple argument that explains their presence.
We will have nothing new to say about this problem in this paper.

{}From the theory of non-linear realisations (NR)
we know that, although the branes
fill out multiplets realising all target space
symmetries linearly, on the branes
the unbroken symmetries are linearly realised while the broken ones are
realised non-linearly. In the context of open string theory one knows
that the supersymmetric field theory on D$p$-branes involve vector multiplets
and are highly non-linear Born--Infeld type theories. Using duality
arguments similar non-linearities can be seen to arise for T5-branes
containing self-dual third rank tensors in $d=p+1=6$ brane dimensions
[\HSWFivebrane,\Fivebrane].

Bagger and Galperin [\Bagger] have recently verified
that the theory of non-linear
realisations applied to supermanifolds embedded into target supermanifolds with
twice the number of anticommuting coordinates
naturally leads to Born--Infeld actions
if vector multiplets are involved. This provides a very nice explanation
for the rather strange form of the Born--Infeld action as being a direct
consequence of the non-linearly realised broken supersymmetries. In this
formalism one introduces derivatives that transform in a well-behaved
manner under the linearly as well as
under the non-linearly realised (super)symmetries. Consistency requirements 
on the constraints imposed on superfields together with  requirement
of symmetry under the linearly as well as the non-linearly realised
supersymmetry imply the Born--Infeld
non-linear action in the case the supermultiplet is chosen to be a 
Maxwell multiplet in 4 dimensions.

Another recently developed approach giving similar results is
the ``doubly supersymmetric geometrical approach'' [\PST,\Sezgin]
or the ``embedding formalism'' [\HSW].
In the latter approach one starts from the torsion
tensor in target superspace and considers the equation that arises when
pulling it back to the super-world-volume.
By introducing a particular embedding
constraint the torsion pull-back equations can, in the only
case analysed explicitly so far namely the T5-brane in 11 dimensions,
 be seen to give rise to exactly
the same non-linear theory as can be argued for from its relation via duality
to the Born--Infeld action of a D4-brane. However, in this formalism the
non-linearly realised supersymmetry plays no r\^ole whatsoever, and it is not
clear that the non-linearities of the action actually have their origin
in some broken symmetries, although this clearly must be the case [\Kallosh].

It is the purpose of this paper to clearify some aspects of the connection
between these two approaches and demonstrate that also for the T5-brane
the non-linearities
of the action stem from an underlying set of broken symmetries.
In section {\old2} we discuss some basic
properties of superembeddings using as an example some results from
the theory of non-linear realisations as well as from the theory of
superembeddings applied to the T5-brane, with a $(6|8)$ super-world-volume,
embedded into a $(7|16)$ target
superspace. Here the notation $(m|n)$ refers to a superspace with $m$ commuting
and $n$ anticommuting coordinates.
Section {\old3} gives the details of this embedding using the
theory of non-linear realisations along the lines of Bagger and Galperin
[\Bagger].
This formalism turns out to generate a rather complicated equation that the
dimension zero components of the torsion tensor induced on the
super-world-volume
must satisfy. Although this equation can be solved explicitly, further
analysis of the system, e.g. deriving the field equations, seems cumbersome
and is not carried out here. Instead we turn in the following sections
to an analysis
of this T5-brane by means of the embedding formalism. In section {\old4} we
show that the theory of non-linear realisations is just a special case
of the embedding formalism, obtained if certain for this formalism
unconventional choices of intrinsic torsion components
are made. In section {\old5}
we then show that the torsion
pull-back equation can be completely analysed and seen to lead to the
non-linearities characteristic of T5-brane field theories, as
already demonstrated for the T5-brane embedded in 11 dimensions
by Howe, Sezgin and West [\HSWFivebrane].
In a final section we summarise our results
and present the conclusions.

\vfill\eject

\section\Ibrane{Superembeddings}In this section we will consider
superembeddings [\Sezgin,\HSW] from a general point
of view, using some explicit results from subsequent sections to
examplify the ideas but leaving the details of special
applications to the later
sections. The different parametrisations of the embedding matrix
to be used in later sections are introduced, and the geometric
properties of the
embeddings are analysed, eventually leading to the torsion pullback equation,
introduces in ref. [\HSW].

Let us consider an arbitrary embedding
$(\ws,\wm)$\lower3pt\hbox{$\,\stack{f}{\epil}\,$}$(\ts,\tm)$,
where the two supermanifolds
have dimensions $(m|n)$ and $(\um|\un)$ respectively. The signature of the
bosonic metric is arbitrary at the moment but later on we will
restrict ourselves to $(D-1,1)$ signature. We will use standard
notation [\HSW] for the local coordinates of the two supermanifolds, i.e.,
$z^M=(x^m;\th^\m)$ and $Z^\tM=(X^\ttm;\Theta^\tmm)$.

We now introduce the embedding matrix\footnote{*}
{Note the difference in notation
compared to refs. [{\xold9},{\xold4}], where the matrix $\SS\te$ does not
denote the embedding matrix,
the latter being denoted $\SS{E_A}^{\tA}$.} $\te_A{}^{\tA}$,
defined in terms of canonical 1-forms $\theta$ by
\eql{\wtco:=f_*\wco=f^*\tco=\we^A\te_A{}^{\tA}\tE_\tA\punkt}{\Embedddef}
Here, $\we^A$ and $\tE_\tA$ are orthonormal basis vectors on the cotangent
space of the world-volume and the tangent of the target space, respectively.
We refer to Appendix A for more details of notation.
The basis vectors $\te_A:=f_*\we_A$ span the tangent
space of the embedded supermanifold. In order to have a complete basis
for the entire
tangent space of the target space, we may also introduce normal
vectors denoted $\te_{A'}$. We will use an overlined index
representing a composite index for the pair $(A,A')$. We will also
introduce a set of dual basis vectors by
\eq{<\te_\bB,\te^\bA>=\d_\bB{}^\bA\punkt}
With these objects at hand we
have the possibility of splitting the canonical 1-form into
tangential and normal terms,
\eq{\tco=\wco+\nco\equiv\te^A\te_A+\te^{A'}\te_{A'}\punkt}
These 1-forms now serve as projectors of vectors down to the tangent and
normal parts respectively, i.e.,
$X^{\parallel}=\wco(X)$, $X^{\perp}=\nco(X)$.
By introducing a target space Lorentz matrix $u_\bB{}^\tA$ relating the basis
$\tE_\tA$ to a frame connected to the embedded surface, it is convenient
to split the embedding matrix as
\eq{\te_A{}^\tA=\te_A{}^\bB u_\bB{}^\tA\punkt}
Concerning the basis $\te_{A'}$ of normal vectors, the choice is
completely arbitrary and physically irrelevant,
and it will soon be clear that in
explicit parametrisations
we can always choose them to be $\te_{A'}{}^\tA=u_{A'}{}^\tA$,
i.e., as part of a Lorentz matrix.

As a starting point for a general superembedding, the orientation in
target superspace of the super-world-volume tangent space is parametrised
by a point in the super-grassmanian
\eql{
        {\rm SGr}[(m|n);(\um|\un)]:={{\rm OSp}(\um|\un)\/{{\rm OSp}(m|n)
        \times {\rm OSp}(\um-m|\un-n)}}\komma}{\Grassmannian}
i.e., there are $m(\um-m)+n(\un-n)$ bosonic parameters and $m(\un-n)+n(\um-m)$
fermionic ones. One way of representing these degrees of freedom is to
introduce the four fields
\eql{
        \tbl{ m_a{}^{b'}&\leftrightarrow&m(\um-m)\komma\hr
                h_\a{}^{\b'}&\leftrightarrow&n(\un-n)\komma\hr
                \ki_a{}^{\b'}&\leftrightarrow&m(\un-n)\komma\hr
                \te_\a{}^{b'}&\leftrightarrow&n(\um-m)\komma\hr }}{\OneFields}
and locally represent the embedding by
\eql{\te_A{}^\tB=\te_A{}^\bB u_\bB{}^\tB
=\mx{u_a{}^{\ul{a}}+m_a{}^{b'}u_{b'}{}^\ta&\ki_a{}^{\a'}u_{\a'}{}^{\ul{\a}}\\
\te_\a{}^{b'}u_{b'}{}^\tb&u_\a{}^{\ul{\a}}+h_\a{}^{\b'}u_{\b'}{}^{\ul{\a}}\\}
\punkt}{\Embeddingone}

If we put this together with the normal vectors we get
\eq{\te_\bA{}^\bB=\mx{
\mx{\d_a{}^b&m_a{}^{b'}\\0&\d_{a'}{}^{b'}\\}&\mx{0&\ki_a{}^{\b'}\\0&0\\}\\
\mx{0&\te_\a{}^{b'}\\0&0\\}&\mx{\d_\a{}^\b&h_\a{}^{\b'}\\0&\d_{\a'}{}^{\b'}\\}
\\}\komma} with the inverse
\eq{\inte_\bA{}^\bB=\mx{
\mx{\d_a{}^b&-m_a{}^{b'}\\0&\d_{a'}{}^{b'}\\}&\mx{0&-\ki_a{}^{\b'}\\0&0\\}\\
\mx{0&-\te_\a{}^{b'}\\0&0\\}&\mx{\d_\a{}^\b&-h_\a{}^{\b'}\\0&\d_{\a'}{}^{\b'}
\\}\\}\punkt}
We notice that the information of the embedding lies entirely in the
matter fields, and that
$u_\bA{}^\tB$ can be chosen arbitrarily. As we will see
in the case of non-linear realisations in sections {\old3} and {\old4},
they may for example
be chosen to be just $\d_\bA{}^\tB$.

In all applications we will choose the part of the embedding matrix
not containing the fields of (\OneFields), i.e., the $u$'s,
to be part of a Lorentz matrix. This choice is always possible, recalling
that the essential property of the embedding matrix is that it defines
the orientation of the embedded hypersurface, so that different embedding
matrices with identical span of the vectors $\te_A$ represent the same
point in the grassmannian (\Grassmannian), and thus the same embedding.
To put it concretely, this degree of arbitrariness in the embedding matrix
is identified with the invariance of its definition (\Embedddef) under
\eql{\tbl{e^A&\rightarrow& e^B{M_B}^A\komma\\
{\te_A}^\tA&\rightarrow&{(M^{-1})_A}^B{\te_B}^\tA\komma\\}}{\Minvariance}
allowing us to go to a representation (\Embeddingone) with lorentzian $u$'s.

The canonical 1-forms
are now expressed in terms of the matter fields in the following way:
\eq{\tbl{ \ico_\ki&=&\ico_0+m'+\te+\ki+h\komma\\
                \nco_\ki&=&\nco_0-m'-\te-\ki-h\komma\\ } }
and if we define new vielbeins by
$\tE_\bA:=u_\bA{}^\tB\tE_\tB$,
we see that
\eq{\tbl{ m'&=&\tE^am_{a}{}^{b'}\tE_{b'}\komma\hfill\\
                \te&=&\tE^\a\te_\a{}^{b'}\tE_{b'}\komma\hr
                \ki&=&\tE^a\ki_a{}^{\b'}\tE_{\b'}\komma\\
                h&=&\tE^\a h_\a{}^{\b'}\tE_{\b'}\hfill\punkt\\ } }

An example of the present parametrisation of the embedding matrix
is given by the NR case (section {\old4}),
where we work in a supersymmetric supermanifold with
$n=\un/2$. There we will see that the fields of (\OneFields) are simply
\eq{\tbl{ m_a{}^{b'}&=&\idel_a\phi^{b'}\komma\hr
        \te_\a{}^{b'}&=&\idel_\a\phi^{b'}-i(\G^{b'}\psi)_\a\komma\hr
        \ki_a{}^{\b'}&=&\idel_a\psi^{\b'}\komma\hr
        h_\a{}^{\b'}&=&\idel_\a\psi^{\b'}\komma\hr } }
where the bosonic matter fields $\phi^{b'}$ are shifted [\HSW] as
\eq{\phi^{b'}=x^{b'}+\ihalf\th\G^{b'}\psi\punkt}
We also see that on
imposing the embedding condition [\Sezgin,\HSW]
\eql{\te_\a{}^\tb=0\komma}{\Embeddcondition}
(this condition, which is a basic geometric relation reducing the number
of field components in the embedding formalism, will be more closely
examined in section {\old4})
we get a relation [\HSW]
\eq{\psi^{\b'}=-\ifrac{$\SS \um-m$}(\G_{c'})^{\b'\a}\idel_\a\phi^{c'}}
between the bosonic fields
$\phi^{b'}$ and the fermionic $\psi^{\b'}$, which are the matter fields
containing the independent degrees of freedom of the embedding ($(\um-m)$ and
$(\un-n)$ respectively). Of course, since these fields are both superfields, 
they contain in general too many physical degrees of freedom. This problem
will be eliminated by analysing the torsion equation together with the
embedding condition.

Returning to our study of the embedding matrix, we note that with the
above parametrisation it is only lorentzian for all
matter fields equal to zero.
It is easy to convince oneself that the field $m_{a}{}^{b'}$ can always
be rotated away by a (target space) Lorentz tranformation:
\eq{m_a{}^b\tilde{u}_b{}^\tc:=u_a{}^{\tc}+m_a{}^{b'}u_{b'}{}^\tc\komma}
so that the $m(\um-m)$ parameters of the orientation of the
bosonic embedding are absorbed into $\tilde{u}$.
The price to be paid for this change of frame is that the fermions rotate
accordingly, and the lower right hand corner of (\Embeddingone) changes.
Again, it is possible to retain the form
$u_\a{}^{\ul{\a}}+h_\a{}^{\b'}u_{\b'}{}^{\ul{\a}}$ by utilising the invariance
(\Minvariance) with a non-lorentzian matrix $M$.
The embedding matrix then takes the form [\HSW]
\eql{\te_\bA{}^\tA=\mx{
        \mx{m_a{}^bu_b{}^{\ul{a}}\\u_{a'}{}^{\ul{a}}\\}
        &\mx{\ki_a{}^{\a'}u_{\a'}{}^{\ul{\a}}\\0\\}\\
        0&\mx{u_\a{}^{\ul{\a}}+h_\a{}^{\b'}u_{\b'}{}^{\ul{\a}}\\
        u_{\a'}{}^{\ul{\a}}\\}\\
                } \komma}{\Embeddingtwo}
where the $u$'s are again lorentzian (the tilde is dropped). 
This Lorentz matrix should of course not
be identified with the one in (\Embeddingone), neither should the
fields denoted by identical symbols.
We have also
dropped the $\te_\a{}^\ta$ term as it will vanish due to the
embedding condition. The new parametrisation also involves a new choice
of basis for the normal vectors.

Equation (\Embeddingtwo) is the form of the embedding matrix to be used
in the rest of the present
section, and in section {\old5}.
The invariance (\Minvariance) used to move between
the two versions (\Embeddingone) and (\Embeddingtwo) of the embedding matrix
involves a redefinition of the intrinsic vielbeins $e^A$, and we may
expect the torsion tensors in the two versions of the theory to exhibit
differences, which is what we will see in the following sections.
It is striking that the seemingly different theories, from a geometric point
of view, are related by a transformation that modifies the 
intrinsic world-volume
geometry by matter fields. We will not analyse the transformations in detail,
but note that they may be worth further study.

The inverse of the modified
embedding matrix is
\eq{\te_\tA{}^\bA=\mx{
        \mx{u_{\ul{a}}{}^b\inm_b{}^a&,&u_{\ul{a}}{}^{a'}\\}&
        \mx{0&,&-u_{\ul{a}}{}^b\inm_b{}^a\ki_a{}^{\a'}\\}\\
        0&\mx{u_{\ul{\a}}{}^\a&,
        &u_{\ul{\a}}{}^{\a'}-u_{\ul{\a}}{}^{\b}h_\b{}^{\a'}\\}\\
                }\komma}
and the canonical 1-forms therefore take the form
\eq{\tbl{ \ico_\ki&=&\ico_0+\ki+h\komma\\ \nco_\ki&=&\nco_0-\ki-h\komma\\ } }
where
\eq{\tbl{ \ki&=&\tE^a\inm_a{}^b\ki_b{}^{\c'}\tE_{\c'}\komma\\ h&=&\tE^\a
        h_\a{}^{\b'}\tE_{\b'}\punkt\hfill\\ }}
Here one should also
mention that none of the free parameters in
$m_a{}^{b'}$ ends up in $m_a{}^b$; the latter becomes determined
completely in terms of $h$.

This is all we will say at this point
about the parametrisation of the embedding matrix.
We will now discuss the origin of the torsion pull-back equation
and in later sections look at some of its solutions. To facilitate the
understanding of the torsion equation we will
point out some conceptual difficulties that may appear in 
connection with it. One
problem is that when working with an embedding of the
type $(\ws,\wm)$\lower3pt\hbox{$\,\stack{f}{\epil}\,$}$(\ts,\tm)$ we have to
consider two different metrics on $\ws$: on the
one hand the a priori (intrinsic) metric on the world-volume $\wm$
and on the other the metric induced by the embedding,
$\im=f^*\tm$. The problem is that we no longer have
one connection on $\ws$ but two, each compatible with one distinct metric.
We will use the notation
$\wdel$ and $\idel$, schematically fulfilling
\eq{\tbl{ \wdel\wm&=&0\komma\\ \idel\im&=&0\punkt\\ }}
Another upcoming
problem is connected to the fact that the embedding is not Lorentz,
unless the matter fields vanish. In order to distinguish the situations,
we will denote the matter fields collectively by $\chi$, and let a 
lorentzian embedding correspond to $\chi\rightarrow0$.

In deriving the torsion equation, we start from the
\Gauss\ equations\footnote\dagger{Ref. [{\xold9}] gives similar equations,
that in addition to our terms on the right hand side also contains
the entities $\SS{\cal L}$, $\SS{\cal L}'$ which will soon be defined. The
difference, as will be clear from the following discussion, resides
entirely in the use of induced contra intrinsic connection in the
derivative.} [\Kobayashi]
\eq{\tbl{ \tdel_XY&=&\idel_XY+\cec'(X,Y)\komma\\
        \tdel_XY'&=&\ndel_XY'+\cec(X,Y')\komma\\ } }
where we have
used the notation $X$ for tangential vectors and $X'$
for normal vectors. We see from these equations that the
covariant derivative splits into a tangential
derivative, a normal derivative and two tensors which
are the so called extrinsic curvatures of the
embedding, also known as the second fundamental
form\footnote*{The term is reserved for $\SS\cec'$, but
$\SS\cec$ is determined from it by
$\SS\tm(\cec'(X,Y),Z')+\tm(X,\cec(Y,Z'))=0$.}.
These equations are purely
tensorial and independent of the form of the embedding. They are also
independent of the intrinsic metric $\wm$ on the world-volume.  If we now
set $Y=\tE_A$ we get
\eq{\tdel\tE_\bA=:\tw_\bA{}^\bB\tE_\bB=\mx{\iw_A{}^B&\ec_A{}^{B'}\\
\ec_{A'}{}^B&\nw_{A'}{}^{B'}}\mx{\iE_B\\\nE_{B'}}\punkt}
The reason for taking $E_\bA$ here instead of $\te_\bA$ is that we need
to make a distinction between whether the embedding is Lorentz or
not. If the embedding is Lorentz then all quantities in these equations
will lie in the algebra {\it spin}($\um$) but not otherwise.
We will therefore make a distinction between the extrinsic
curvatures of the two types of embeddings by denoting the extrinsic
curvature of a Lorentz embedding by roman letters and a matter
triggered embedding by calligraphic ones. From the \Gauss\ equations
it follows that
\eq{\cec_{AB}{}^{C'}=<\tdel_A(\te_B{}),\te^{C'}> \punkt}
This means
that, as the $\te_\bA$ tend to $\tE_\bA$ as $\ki\rightarrow0$, the
extrinsic curvature will tend to the Lorentz one, i.e.,
$\cec_{AB}{}^{C'}|_{\ki=0}=\ec_{AB}{}^{C'}$.
Of course
\eq{\ec_{AB}{}^{C'}=<\tdel_A(\tE_B{}),\tE^{C'}>=<\tdel_A(u_B),u^{C'}>
        \komma}
where $u_B={u_A}^\tA E_\tA$ and $u^{A'}=E^\tA{u_\tA}^{A'}$.
Now since we will use the intrinsic 
world-volume metric $\wm$ as an auxiliary field in the
forthcoming torsion equation, we need a
relation between the two connections on $\ws$. Let us define a
difference operator of the two of them by
\eq{\cl:=\idel-\wdel \punkt}
This operator is of course a tensor.
Proceeding as for the extrinsic curvature we let
$\cl|_{\ki=0}=:L$.
We will also extend our covariant
derivatives on $\ws$ to act on world-volume vectors as well as
target space vectors and denote them as $\ioldel$ and $\woldel$. This
enables us to note the following important relations
\eq{\tbl{ \ioldel(\wtco)&=&\cec'\komma\hfill\\
                \woldel(\wtco)&=&\cl+\cec'\\ } }
(these are tensor equations, so there is no wedge product involved),
{}from which we see that the tensors can be written
\eq{\tbl{ \cl_{AB}{}^C&=&\woldel_A(\te_B{}^{\tC})\te_{\tC}{}^C\komma\\
        \cec_{AB}{}^{C'}&=&\woldel_A(\te_B{}^{\tC})\te_{\tC}{}^{C'}\komma\\
                } }
and consequently
\eq{\tbl{ L_{AB}{}^C&=&\woldel_A(u_B{}^{\tC})u_{\tC}{}^C\komma\\
        \ec_{AB}{}^{C'}&=&\woldel_A(u_B{}^{\tC})u_{\tC}{}^{C'}\punkt\\
                } }
Let us introduce yet another covariant derivative
in order to get relations between these fields:
\eq{\woldelmod=\woldel|_{\hbox{\fiverm diag}}+\hat{X}\komma}
where
\eq{\hat{X}_\bA{}^\bB:=\mx{L_A{}^B&0\\0&L_{A'}{}^{B'}\\}\komma}
and where the connection in the first term on the right hand side contains
the target space connection
projected on the part not mixing tangential and normal directions.
Let us also define
\eq{{\cal X}_\bA{}^\bB:=\woldel(\te_\bA{}^\tC)\te_\tC{}^\bB
        \equiv\mx{\cl_A{}^B&\cec_A{}^{B'}\\\cec_{A'}{}^B&\cl_{A'}{}^{B'}\\}
        \punkt}
This notion is natural because it will tend to $L$ and $\ec$ as
$\ki\rightarrow0$. We now get the relation between the fields
\eq{{\cal X}_\bA{}^\bB=\hat{X}_\bA{}^\bB+
        \woldelmod(\te_\bA{}^\bC)\inte_\bC{}^\bB+\te_\bA{}^C\ec_C{}^{D'}
        \inte_{D'}{}^\bB+
        \te_\bA{}^{C'}\ec_{C'}{}^D\inte_D{}^\bB \komma}
{}from which,
if we look at our parametrisation of the embedding
matrix in particular, we now get the following relations
\eql{\tbl{ {\cal L}_b{}^{c}&=&L_b{}^{c}+(\woldelmod
                m_b{}^d)\inm_d{}^{c}\komma\hfill\\
        {\cal L}_b{}^{\c}&=&\ki_b{}^{\b'}\ec_{\b'}{}^{\c}\komma\hfill\\
        {\cal L}_\b{}^{c}&=&0\komma\hfill\\
        {\cal L}_\b{}^{\c}&=&L_\b{}^{\c}+h_\b{}^{\b'}\ec_{\b'}{}^\c
                \komma\hfill\\
        {\cal K}_b{}^{c'}&=&m_b{}^c\ec_c{}^{c'}\komma\hfill\\
        {\cal K}_b{}^{\c'}&=&\woldelmod\ki_b{}^{\c'}-(\woldelmod
                m_b{}^c)\inm_c{}^{d}\ki_d{}^{\c'}
                -\ki_b{}^{\b'}\ec_{\b'}{}^{\c}h_\c{}^{\c'}\komma\hr
        {\cal K}_\b{}^{c'}&=&0\komma\hfill\\
        {\cal K}_\b{}^{\c'}&=&K_\b{}^{\c'}+\woldelmod
                h_\b{}^{\c'}-h_\b{}^{\b'}\ec_{\b'}{}^\c
                h_\c{}^{\c'}\komma\hfill\\
        {\cal K}_{b'}{}^{c}&=&K_{b'}{}^{d}\inm_d{}^c\komma\hfill\\
        {\cal K}_{b'}{}^{\c}&=&0\komma\hfill\\
        {\cal K}_{\b'}{}^{c}&=&0\komma\hfill\\
        {\cal K}_{\b'}{}^{\c}&=&K_{\b'}{}^{\c}\komma\hfill\\
        {\cal L}_{b'}{}^{c'}&=&L_{b'}{}^{c'}\komma\hfill\\
        {\cal L}_{b'}{}^{\c'}&=&-K_{b'}{}^{d}\inm_d{}^e\ki_e{}^{\c'}
                \komma\hfill\\
        {\cal L}_{\b'}{}^{c'}&=&0\komma\hfill\\
        {\cal L}_{\b'}{}^{\c'}&=&L_{\b'}{}^{\c'}-K_{\b'}{}^{\c}h_\c{}^{\c'}
                \punkt\hfill\\} }{\clkrel}
Some of the zeroes are directly related to the embedding condition
(\Embeddcondition). The virtue of these relations is that they display
explicitly which properties of the
geometry are induced by matter fields.
They are important because we will use them in the process of
solving the torsion equation. We now turn to the issue of deriving
the torsion equation, which is the 
final subject of this section.

If we look
at the \Gauss\ equations we see that
\eq{\tmT(X,Y):=\tdel_XY-\tdel_YX-[X,Y]=\iT(X,Y)+\eT'(X,Y)\komma}
where $\iT(X,Y)$ is the induced torsion inherited from the connection on
$T\ts$ and
\eq{\eT'(X,Y):=\cec'(X,Y)-\cec'(Y,X) }
is called the extrinsic
torsion of the embedding. But we know that we have a relation between
the induced torsion and the intrinsic torsion denoted $\wT$ from
the relation of the two connections on $\ws$. This relation is
\eq{\iT(X,Y)=\wT(X,Y)+\cl(X,Y)-\cl(Y,X) \komma}
which together with the relation
\eq{\woldel\wedge\wtco=\wedge\cl+\eT'=-\wT+\iT+\eT' }
(the notation $\wedge\cl$ meaning the antisymmetric part)
finally
yields the torsion equation in the form
\eq{\woldel\wedge\wtco(X,Y)+\wT(X,Y)=\tmT(X,Y) \komma}
where of course
$X,Y$ everywhere are super-world-volume tangent vectors.
This is nothing but the usual
torsion equation that figures in the physics literature
[\Sezgin,\HSW].
Putting
$X=\te_A$ and $Y=\te_B$ and contracting with $\tE^\tC$ we get it in the more
transparent form
\eql{\woldel_A\te_B{}^\tC-(-1)^{AB}\woldel_B\te_A{}^\tC
        +\wT_{AB}{}^C\te_{C}{}^\tC
        =(-)^{A(B+\tB)}\te_{B}{}^\tB\te_A{}^\tA\tT_{\tA\tB}{}^\tC \punkt}
        {\teq}
In order to solve this equation we will project it onto the
tangent and the normal directions, respectively, giving
\eq{2\cl_{[AB)}{}^C+\wT_{AB}{}^C=\iT_{AB}{}^C } and
\eq{2\cec_{[AB)}{}^{C'}=\iT_{AB}{}^{C'}\komma} where
the graded anti-symmetrisation is defined by
$V_{[AB)}:={1\/2}(V_{AB}-(-1)^{AB}V_{BA})$.
Now for the parametrisation in eq. (\Embeddingtwo)
we have the following induced
torsion components (if the target space is flat):
\eql{\tbl{ \iT_{ab}{}^c&=&i[\ki_a{}(\G^d)\ki_b]\inm_d{}^c\komma\hr
        \iT_{ab}{}^\c&=&0\komma\hr
        \iT_{a\b}{}^c&=&-i[\ki_a(\G^d)h_\b]\inm_d{}^c\komma\hr
        \iT_{a\b}{}^\c&=&0\komma\hr
        \iT_{\a\b}{}^c&=&-i[(\G^d)_{\a\b}+h_\a(\G^d)h_\b]\inm_d{}^c\komma\hr
        \iT_{\a\b}{}^\c&=&0\komma\hr \iT_{ab}{}^{c'}&=&0\komma\hr
        \iT_{ab}{}^{\c'}&=&-i[\ki_a{}(\G^d)\ki_b]\inm_d{}^e\ki_e{}^{\c'}
                \komma\hr
        \iT_{a\b}{}^{c'}&=&-i\ki_a(\G^{c'})_{\b}\komma\hr
        \iT_{a\b}{}^{\c'}&=&i[\ki_a(\G^d)h_\b]\inm_d{}^e\ki_e{}^{\c'}\komma\hr
        \iT_{\a\b}{}^{c'}&=&-i2h_{(\a}(\G^{c'})_{\b)}\komma\hr
        \iT_{\a\b}{}^{\c'}&=&i[(\G^d)_{\a\b}+h_\a(\G^d)h_\b]
                \inm_d{}^e\ki_e{}^{\c'}\punkt\hr} }{\trel}
The $\Gamma$ matrices have been split according to appendix B,
and summed $\a'$ indices are suppressed, e.g. $h_\a(\G^d)h_\b\equiv 
{h_\a}^{\a'}(\bar\G^d)_{\a'\b'}{h_\b}^{\b'}$.
Together with the expressions for the fields
$\cec$, $\cl$ and of course $\wT$ it is just to begin
solving for the matter fields. We already here see
that the solutions will depend on the chosen
intrinsic world-volume torsion $\wT$, but we will come back to
this in later sections.  If we instead look at the
case of our first parametrisation, given in eq. (\Embeddingone), where we had a
direct coupling to the NR case, we get
\eql{\tbl{ \iT_{ab}{}^c&=&i\ki_a{}(\G^c)\ki_b\komma\hr
                \iT_{ab}{}^\c&=&0\komma\hr
                \iT_{a\b}{}^c&=&-i\ki_a(\G^c)h_\b\komma\hr
                \iT_{a\b}{}^\c&=&0\komma\hr
                \iT_{\a\b}{}^c&=&-i[(\G^c)_{\a\b}+h_\a(\G^c)h_\b]\komma\hr
                \iT_{\a\b}{}^\c&=&0\hr } }{\trelii}
(again, although the fields denoted by the same letters in (\trel)
and (\trelii) are related by field redefinitions, they
should by no means be identified), which we will see in later
sections is nothing but the relations for the torsion
derived from the algebra of the induced covariant
derivatives.

\section\bag{The $D=6$ tensor multiplet and non-linear realisations}In this
section we will review the basic steps of the theory of non-linear realisations
[\Volkov], which is a systematic way of studying the properties
of Goldstone fields. It is well-known that the spontaneous breaking of
supersymmetry gives rise to a massless spin-${1\/2}$ Goldstone
fermion [\Akulov]. This fermion then belongs to the massless
multiplet of the residual unbroken supersymmetry.  However, the choice
of the Goldstone multiplet is not unique.  The partial breaking of
$N=2$ supersymmetry to $N=1$ in four dimensions was studied in
[\Bagger], for three different multiplets.  We will use non-linear
realisations to describe the spontaneous breaking of $N=1$
supersymmetry in $D=7$ to $N=(1,0)$ in $D=6$ 
and pick the self-dual tensor multiplet in
6 dimensions [\Strathdee,\HST] as the Goldstone multiplet.

Let $ \underline{{\cal M}}^{(7|16)}$ be a flat $N=1$ target
superspace with local coordinates
$Z^{\tM}=(X^{\underline{m}},\Theta^{\tmm})$.
Our starting point is the 7-dimensional $N=1$ supersymmetry
algebra
\eq{ \{ Q_{\taa}, Q_{\tbb}\}=(\Gamma^{\ta})_{\underline{\a\b}}P_{\ta}\punkt  }
Making the 7$\rightarrow$6+1 split, using the conventions of appendix B, this
algebra reads:

\eq{
        \tbl{ \{ Q_{\a}^i, Q_{\b}^j \}&=&
                {}\e^{ij}(\g^a)_{\a\b}P_a\komma\hfill\\ \{Q_{\a}^i ,
                S^{\b}_j
                \}&=&\delta_{\a}{}^{\b}\delta^{i}{}_{j}Z\komma\hfill \\
                \{S^{\a}_i, S^{\b}_j \}&=&
                \e_{ij}(\g^a)^{\a\b}P_a\punkt\hfill } } 
Here $\e^{ij}$ is
the invariant tensor of the SU(2) automorphism group.
{}From a 6-dimensional point of view, this is
an $N=(1,1)$ algebra with a central charge $Z$, the
momentum in the seventh direction. We now consider the
partial breaking of this $N=(1,1)$ algebra down to
$N=(1,0)$.
Let ${Q_{\a}^i}$ be the unbroken $N=1$ supersymmetry generator and
${S^{\a}_i}$ its broken counterpart. A parametrisation of
the $N=1$ target superspace $ \underline{{\cal M}}^{(7|16)}$ suitable
for our problem is
\eq{
        \W=\exp[i(x^aP_a+\th^\a_iQ^i_\a)]\exp[i(yZ+\psi_\a^iS_i^\a)]\punkt}
Now the spinor $\psi_\a^i=\psi_\a^i(x,\th)$ is the Goldstone
superfield associated with the broken generator ${S^{\a}_i}$, and the
scalar $y=y(x,\th)$ is the Goldstone superfield associated with the
central charge $Z$. Here we have employed the "static gauge" for the
splitting of target space coordinates:
\eql{\matrix{X^m&=&x^m\komma&&&&X^{6}&=&y(x,\theta)\komma\cr
          \Theta^\mu&=&\theta_i^\mu\komma &&&& \Theta^{{\mu}'}
          &=&\psi_{\mu}^i(x,\theta)\punkt\cr}}{\Coorddef}
Note that this construction
naturally corresponds to the embedding
${\cal M}^{(6|8)}\hookrightarrow\underline{{\cal M}}^{(7|16)}$,
where the Goldstone fields are bosonic and fermionic coordinates
describing the shape of the supersurface ${\cal M}^{(6|8)}$, which
automatically breaks half of the supersymmetry.

The $S$-supersymmetry acts with $g$=exp$(i\eta S)$ on $\Omega$ by
left multiplication, $g\Omega=\Omega'$, which induces a
transformation on the bosonic coordinates
\eq{ \delta_{\eta}x^a=-\ihalf\eta{}\bl{\g}^a\psi\punkt}
This in turn makes the transformations of the Goldstone fields
contain non-linear terms, in addition
to the usual shifts:
\eq{
            \tbl{ \delta_{\eta}\psi_{\a}^i & = &\eta_{\a}^i
                +\ihalf\eta{}\bl{\g}^a\psi\partial_a\psi_{\a}^i \komma\hfill\\
                 \hfill \delta_{\eta}y & = &-{i\/2}\eta{}\theta
                +\ihalf\eta{}\bl{\g}^a\psi\partial_ay\punkt\hfill
                 } }

Since the Cartan 1-form $\W^{-1}d\W$ takes its value in the
supersymmetry algebra, we can para\-metrise it in the following way
\eq{
        \W^{-1}d\W=i[E^aP_a+E^6Z+E^\a_iQ_\a^i+E_\a^iS^\a_i]\punkt  }
This expansion gives the covariant world-volume Goldstone 1-forms:
\eql{\matrix{E^a&=&dx^a-{i\/2}[d\th\tilde\g^a\th+d\psi\bl{\g}^a\psi]\komma&
          &&&&E^\a_i&=&d\th^\a_i\komma\\
          E^6&=&dy-{i\/2}[d\th\psi+d\psi\th]\komma\hfill&&&&&
          E_\a^i&=&d\psi_\a^i\punkt}}{\GOneform}
Here we use the notation
$\tilde\g^a:=\epsilon^{ij}(\g^a)_{\a\b}$ and
$\bl\g^a:=\epsilon_{ij}(\g^a)^{\a\b}$. The vielbein matrix
${E_{M}}^A$ is found from the expansion of the world-volume
1-form $E^A=(E^a,E^\a_i)$ with respect to the coordinate
differential $dz^M=(dx^m,d\th_{i}^\mu)$ of the world-volume,
$E^A=dz^M{E_{M}}^A$.
The $N=2$ derivatives induced by the
Goldstone superfields are then given by\footnote\dagger{These 
induced covariant
derivatives, denoted $\SS\idel$ in the present paper (see appendix A) equal
those denoted $\SS\wdel$ in ref. [{\xold6}].}
\eq{
             {\idel}_A =(E^{-1})_A{}^M \partial_M\punkt  }
These covariant derivatives can be explicitly written as:
\eq{
            \tbl{ {\idel}_a&=&(E^{-1})_a{}^m\partial_m \komma\hfill\\
                   {\idel}_\a^i & = &
                 D_{\a}^i+{i\/2}(D_{\a}^i\psi)\bl{\g}^a\psi\idel_a
        \punkt\vbox to 13pt{\vfill}\hfill } }
It is interesting to note that the covariant derivative ${\idel}_a$
satisfies the implicit relation
\eq{{\idel}_a  =  D_a+ \ihalf(D_a\psi)\bl{\g}^a\psi\idel_a\komma }
which simply follows from solving for $\partial_m$ above. This expression then
most easily gives the expression for $ {\idel}_\a^i$ above,
which otherwise, when directly solved for as the dual of (\GOneform), is
expressed in terms of the bare derivatives $\partial_\a$. 
Here $D_{\a}^i$ and $D_a$ are the ordinary flat $N=1$ covariant
derivatives.  It is then straightforward to calculate the algebra of
the $N=2$ covariant derivatives [\Bagger]:
\eq{
    \tbl{ [\idel_a,\idel_b]&=& -i(\idel_a\psi)\bl{\g}^c(\idel_b\psi)
\idel_c\komma\hfill\\ 
[\idel_a,\idel_\a^i]&=
&i(\idel_\a^i\psi)\bl{\g}^b(\idel_a\psi) \idel_b\komma
	\vbox to 13pt{\vfill}\hfill\\
\{ \idel_\a^i,\idel_\b^j \}&=&i\epsilon^{ij}\g_{\a\b}^a\idel_a
+i(\idel_\a^i\psi)\bl{\g}^a(\idel_\b^j\psi)\idel_a\komma 
	\vbox to 13pt{\vfill} } }
in accordance with eq. (\trelii).

It is convenient to introduce the scalar superfield
\eq{\Phi:=\half\theta\psi-iy\komma}
which, in particular, implies:
\eq{E^6 = id\Phi-id\th\psi\punkt}
This shift, anticipated in section \Ibrane, is necessary in order
to obtain a scalar superfield under the 6-dimensional supersymmetry algebra.
Note that at this stage there is no relation between the Goldstone
fields.  We now impose the irreducibility condition
\eq{E_{\a}^{i{}6}=0\komma} or equivalently
\eq{   \psi_\a^i = \idel_\a^i \Phi\punkt}
In the next section we will see that this constraint is inherent in the
embedding formalism, where it is part of the embedding condition
${\te_\a}^\ba=0$. In the present treatment its remaining components
${\te_\a}^a$ vanish trivially.

The on-shell self-dual tensor
multiplet in 6 dimensions is given by
\eq{ (1,0)\oplus 2(\half,0)\oplus(0,0)\,\leftrightarrow\,
A_{ab}^+\oplus\psi_{\a}^i\oplus\phi\komma}
where $\psi_{\a}^i$ and $\phi$ are the leading components of the
spinor Goldstone superfield and the shifted scalar superfield,
respectively, and where we have used the standard labeling of the
massless particles by the helicity states of the little group
Spin(4)$\,\approx\,$SU(2)$\times$SU(2). The minimal $N=(1,0)$
supersymmetry in
6 dimensions does indeed admit this tensor multiplet [\Strathdee].
Here $A$ is a 2-form potential coming from the
symmetric bispinor superfield [\HST]
\eq{    F_{\a\b} := \half \idel_{(\a i}\idel_{\b)}^{i}\Phi
:= \idel_{\a\b}\Phi\komma}
which corresponds to a self-dual field strength
$F_{\a\b}=\fraction6(\Gamma^{abc})_{\a\b}F_{abc}$.
It has been suggested [\Bagger] that there might be an extension of the
$N=2$ supersymmetry which associates a Goldstone-like symmetry with
this field and the tensor gauge field might itself be a Goldstone
field.

To describe the on-shell self-dual tensor multiplet, the
superfield $\Phi$ has to be further constrained. This constraint is
most easily expressed in terms of the $N=2$ covariant derivatives.  An
appropriate constraint can be found from the decomposition${}^*$ [\HST]
\eq{ \idel_\a^i\idel_\b^j\equiv-\half{T}_{\a\b}^{{ij}{}a}\idel_a
                                +\epsilon^{ij}\idel_{\a\b}
                        +\idel_{[\a}^{(i}\idel_{\b]}^{j)}\punkt}
Let us first consider the linear
case. This decomposition then reads
\def\footlineone{\vtop{\baselineskip=9pt
        \hrule width.6\hsize\hfill\break
        \indent ${}^*$ \vtop{\hsize=14cm\noindent\xrm 
We label the irreducible parts
of the decomposition as
$\SS( {\bf 4},{\bf 2})\otimes( {\bf 4},{\bf 2})\equiv ({\bf 6},{\bf 1})\oplus
({\bf 10},{\bf 1})\oplus({\bf 6},{\bf 3})\oplus({\bf 10},{\bf 3})$,
reflecting the
group structure Spin(1,5)$\SS\times$SU(2).
	}}
        \vskip-.7\baselineskip}
        \foottest=1
\eq{D_\a^i D_\b^j\equiv\ihalf\epsilon^{ij}(\gamma^a)_{\a\b}\partial_a
+\epsilon^{ij}D_{\a\b}+
D_{[\a}^{(i}D_{\b]}^{j)}\komma} since the representation $({\bf 10},{\bf
3})$ vanishes,
$D_{(\a}^{(i}D_{\b)}^{j)}\equiv 0$.
It is easily shown [\HST] that the constraint
\eq{D_{[\a}^{(i}D_{\b]}^{j)}\Phi=0\komma}
postulating the absence of fields in the representation $({\bf 6},{\bf 3})$,
describes the on-shell self-dual tensor multiplet.

Turning to the full
non-linear case again, we make the assumption, later to be verified,
that the constraint generalises as
\eql{ \idel_{[\a}^{(i}\idel_{\b]}^{j)}\Phi= 0\punkt}{\NLconstraint}
The world-volume torsion is given by the implicit equation
\eq{  \{ \idel_\a^i,\idel_\b^j \}
        =:-T_{\a\b}^{ij\,a}\idel_a =
     i\epsilon^{ij}(\g^a)_{\a\b}\idel_a
        +i(\idel_\a^i\psi)\bl{\g}^a(\idel_\b^j\psi)\idel_a\punkt  }
Note that this is a highly non-linear equation, since the fact that
$\psi_\a^i = \idel_\a^i \Phi$ implies that also the right hand side
contains torsion. We now
proceed to give an explicit expression for this
component of the induced torsion on-shell.  Using the
constraint above and acting on the scalar superfield
$\Phi$, we get the torsion equation on the form
\eql{ 2{T}_{\a\b}^{ij}= \g_{\a\b}^{ij}+
                ({T}_{\a\g}^{ik}+\epsilon^{ik}F_{\a\g})\g_{kl}^{\g\d}
                ({T}_{\b\d}^{jl}+\epsilon^{jl}F_{\b\d})\komma}
	{\ImplicitTorsion}
where ${{T}_{\a\b}^{ij}:=-{1\/2}{T}_{\a\b}^{ij\,a}\idel_a\Phi}$ and
${\g_{\a\b}^{ij}:=i\epsilon^{ij}(\g^a)_{\a\b}\idel_a\Phi}$.
The crucial point is that the
totally symmetric representation
$(\bf{10},\bf{3})$ drops out of the torsion
after the on-shell constraint is imposed, and
therefore
\eq {{T}_{\a\b}^{ija}=\epsilon^{ij}{{T}_{\a\b}}^{a}\punkt}

The torsion equation can then be written as the matrix equation
\eq{ 2T=\g+(T+F)\g(F-T)\komma}
by extracting an overall $ \epsilon^{ij}$. It is convenient to
introduce a matrix $A$ such that $A^2:=\g$. Then let $B:=AFA$ and
$X:=ATA$. The torsion equation now reads
\eq{2X=A^4+(X+B)(B-X)\komma}
with the solution
\eq{X=-1+\sqrt{(1+A^4+B^2)}\punkt}
Note that $A^4=(\idel\Phi)^2$. In the weak-field expansion we
get
\eq{X=\sum_{n=1}^\infty\left(\matrix{{1\/2}\cr n\cr}\right)A^{4n}
        +\sum_{n=1}^\infty
      \left(\matrix{{1\/2}\cr n\cr}\right)(1+A^4)^{{1\/2}-2n}B^{2n}\punkt}
The torsion is then explicitly given by
\eql{ T=\half\g
+\sum_{n=1}^\infty\left(\matrix{{1\/2}\cr n+1\cr}\right)(\idel\Phi)^2\g+
\sum_{n=1}^\infty\left(\matrix{{1\/2}\cr n\cr}\right)
(1+(\idel\Phi)^2)^{2n-1}F(\g F)^{2n-1}\punkt}{\ExplicitTorsion}
It is essential for obtaining ${T_{\a\b}}^a$
that it is possible to extract a factor
$\idel\Phi$. To check that the supersymmetry algebra closes
on the self-dual tensor multiplet it is sufficient to calculate
$\idel_\a\idel_\b\idel_\g\Phi$. This check is cumbersome due to the fact
that ${T_{a\a}}^b$ is given by a linear equation which in turn depends on 
${T_{\a\b}}^a$.
{}From the solution (\ExplicitTorsion), we see that the
supersymmetry transformations of the component fields will be
extremely non-linear. However, no more fields are
generated. Hence our $N=2$ covariant constraint, eq. (\NLconstraint)
is correct, and puts the theory on-shell.

We conclude that the $N=(1,0)$ self-dual
tensor multiplet in 6 dimensions can indeed be given an interpretation as
a Goldstone multiplet for the chirally broken $N=(1,1)$ (or, actually
7-dimensional) supersymmetry, which is natural from a brane
viewpoint. Since there is no lagrangian formulation of the theory
(without the introduction of auxiliary fields [\SDLagrangian], which however do
not seem to have any natural interpretation in the present framework),
the program pursued for e.g. the Maxwell multiplet in ref. [\Bagger],
where a lagrangian formulation was derived,
has no counterpart for this supermultiplet.
The constraint (\NLconstraint), which is the most naive covariantisation
of the irreducibility constraint of the linear theory, turns out
to be consistent, and encodes the full non-linear field equations.
Due to the complicated nature of the torsion, given by an implicit
relation (\ImplicitTorsion) solved as (\ExplicitTorsion), the derivation
of the field equations for the component fields becomes cumbersome,
and will not be performed here.
We note that the explicit form of the torsion may be summarised as
a formal square root, an observation that probably is connected
to the relation with Born--Infeld theory.

\section\bag{Non-linear realisations in the embedding formalism}In this
section we review some of the salient features of the
embedding formalism, as applied to the superembedding of the 5-brane
in $D=7$. The ``embedding formalism'' [\HSW] or the ``doubly supersymmetric
approach'' [\PST,\Sezgin] to describe $p$-brane dynamics\footnote\dagger{We 
do not
strictly want to call these separate formalisms; rather we would like to
reserve the former term for the specific procedure of extracting information
about the dynamics from the torsion equation.} are based on a
geometrical condition specifying the superembedding of a world-volume
into target space. This condition can furthermore be obtained from a
"generalised geometrical action principle" [\PST]. The power of the
formalism was demonstrated in [\HSWFivebrane] for the T5-brane in 
11 dimensions,
where the embedding condition was postulated and
supersymmetric equations of motion obtained before a complete
supersymmetric action for them was constructed [\Fivebrane].

Consider the flat target superspace $ \underline{{\cal M}}^{(7|16)}$
locally parametrised with coordinates
$Z^\tM=(X^{\underline{m}},\Theta^\tmm)$,
and introduce the supersymmetric cotangent basis 1-forms in target
space
\eq{
     \tbl{ \Pi^{\ttm} &= & dX^{\ttm} - {i\/2} d \Theta
               \Gamma^{\ttm}\Theta\komma\hfill\\ \Xi^{\tmm} &= & d
               \Theta^{\tmm}\punkt\hfill } }
An arbitrary frame is obtained by SO(1,6) rotations
\eq{     \tbl{    E^{\ta}     &= & \Pi^{\ttm} u_{\ttm}{}^{\ta}
        &=& dZ^{\tM} E_{\tM}{}^{\ta}\komma\hfill\\
                  E^{\taa} &= & \Xi^{\tmm} u_{\tmm}{}^{\taa} &= &
                                    dZ^{\tM} E_{\tM}{}^{\taa}\punkt\hfill }
                                    }
Here $u_{\ttm}{}^{\ta}$ and
$u_{\tmm}{}^{\taa}$ are the
``Lorentz harmonics''.  The embedding
matrix $\te_A{}^\tA$ is defined as
the pullback of the target space
1-form $E^\tA$ onto the
world-volume:
\eq{   \te_A{}^\tA :=E_A (f^*E^\tA) = E_A{}^M(\partial_MZ^\tM)E_\tM{}^\tA =
( \idel_A Z^\tM )E_\tM{}^\tA\punkt } Here $\idel_A$ is the induced
covariant derivative on the world-volume.  The essential ingredient of
the doubly supersymmetric approach is the "geometro-dynamical
condition" [\PST,\Sezgin], or the embedding condition [\HSW]
\eq{ \te_\alpha{}^{\ta}=0\punkt }
Geometrically, this is simply the requirement that, at any point of
 ${\scr M}$, the odd tangent space to ${\scr M}$ lies entirely within
 the odd tangent space to $\underline{{\scr M}}$. In a number of
 interesting cases [\HSW],
the integrability condition for this constraint is so strong that it
 reproduces all the equations of motion for the extended object. 
This happens e.g. for the T5-brane in $D=11$ [\HSW,\HSWFivebrane]. 
In the next section, however, we show that the embedding condition alone
 is not sufficient to put the $D=7$ 5-brane multiplet on-shell. It
 has to be augmented by a suitable constraint, as conjectured in ref.
[\HSWFivebrane].

The embedding matrices can be read off from the induced vielbeins on
the world volume. Expressed in terms of the Goldstone fields, they
are, as mentioned in section \Ibrane:
\eq{
     \tbl{ \te_a{}^\ta &=&\delta_a{}^{\ta}
        +i(\idel_a\Phi)\delta_{6}{}^{\ta}\komma\hfill\\ \te_{\a}{}^{\taa}
           &=&\delta_{\a}{}^{\taa}
        +(\idel_{\a}\Theta^{\a'})\delta_{\a'}{}^{\taa}\komma\hfill } } and
\eq{ \te_a{}^{\taa}=(\idel_a\Theta^{\a'})\delta_{\a'}{}^{\taa}\punkt }
The embedding condition reads explicitly
\eq{\te_{\a}^{i\ta}=\idel_{\a}^i X^{\ta}-\ihalf(\idel_{\a}^i\Theta)
\Gamma^{\ta}\Theta=0\punkt}
In particular, $\te_{\a}^{i6}=0$ gives the $D=7$ non-linear "master
constraint" of [\HSW]:
\eq{\psi_\a^i=\idel_\a^i\Phi\komma }
($\psi$ being the normal spinor coordinate as in eq. (\Coorddef))
as advertised in section \Ibrane.
We know that the linearised version of the above
constraint is not sufficient to put our theory on-shell. In the next
section we show that this is also true at the non-linear level, without
using a particular gauge, e.g. the static gauge.

Turning now to the induced world-volume torsion, it can be calculated
{}from the integrability condition for the embedding matrix,
$\idel_{(\a}\te_{\b)}{}^{\ta}=0$,
which gives
\eq{ i{T}_{\a\b}{}^c \te_c{}^\tc
=\te_\a{}^\taa \te_\b{}^\tbb(\Gamma^\tc)_{\underline{\alpha\beta}}
     \punkt}
This is also known as the "twistor constraint" since
$\te_{\a}{}^{\taa}=\idel_{\a}^i\Theta^{\taa}$
is a twistor-like bosonic superfield.  The world-volume torsion is then
given by the equation
\eq{ i {T}_{\a\b}^{{ij}\,a} =
     \epsilon^{ij}(\g^a)_{\a\b} +\epsilon_{kl}
        (\idel_\a^i\idel_\g^k\Phi)\g^a(\idel_\b^j\idel_\d^l\Phi)}
(see eq. (\trelii)),
which is identical to the one obtained in the
non-linear realisation formalism.

\vfill\eject
\section\EqOM{The equations of motion}We are now going to
derive the equations of motion for the super-5-brane in
$7$ dimensions. As target space we will choose a flat $D=7$ superspace,
i.e., all torsion components vanish except for
\eq{
\tT_{\ul{\a\b}}{}^\tc=-i(\G^\tc)_{\ul{\a\b}}\punkt
}
The intrinsic world-volume geometry is chosen to be $N=1$, $d=6$ conformal
supergravity [\Gates] and the constraints that we will need in order
to obtain the
equations of motion are
\eq{
\wT_{\a\b}{}^c=-i(\G^c)_{\a\b}
}
and
\eq{
\wT_{\a\b}{}^\c=\wT_{\a b}{}^c=\wT_{ab}{}^c=0\punkt
}

The fields occurring in the following equations are those found in the
parametrisation (\Embeddingtwo) of the embedding matrix.
We start by extracting the information hidden in (\teq) using the constraints
(\clkrel) and (\trel). We thus obtain
\eq{\tbl{
        \hfill\hbox{(i)}&\hfill \woldelmod_{[a}m_{b]}{}^c&=&
                \ihalf(\chi_a \G^c\chi_b)-L_{[ab]}{}^d m_d{}^c
                \komma\hr
        \hfill\hbox{(ii)}&\hfill m_{[a}{}^d K_{b]d}{}^{c'} &=&0\komma\hr
        \hfill\hbox{(iii)}&\hfill\wT_{ab}{}^\c &=
                & 2\chi_{[a}{}^{\b'} K_{b]\b'}{}^\c\komma\hr
        \hfill\hbox{(iv)}&\hfill\woldelmod_{[a} \chi_{b]}{}^{\c'} &=&
                -\chi_{[a}{}^{\b'}K_{b]\b'}{}^\c h_{\c}{}^{\c'}-L_{[ab]}{}^c
                \chi_c{}^{\c'}\komma\hr
        \hfill\hbox{(v)}&\hfill\woldelmod_\b m_a{}^c &=
                & i(\chi_a \G^c h_\b)-L_{\b a}{}^d\inm_d{}^c\komma\hr
        \hfill\hbox{(vi)}&\hfill m_a{}^d K_{\b d}{}^{c'} &=
                & i(\chi_a \G^{c'})_\b \komma\hr
        \hfill\hbox{(vii)}&\hfill\wT_{a\b}{}^\c &=
                & -\chi_a{}^{\b'}K_{\b\b'}{}^\c-h_\b{}^{\b'}K_{a\b'}{}^\c
                - L_{a\b}{}^\c\komma\hr
        \hfill\hbox{(viii)}&\hfill\woldelmod_a h_{\b}{}^{\c'}
                -\woldelmod_\b \chi_a{}^{\c'} &=& L_{\b a}{}^d \chi_d{}^{\c'}
                + h_\b{}^{\b'}K_{a\b'}{}^\c h_\c{}^{\c'}\hr
        &&&     \quad+ \chi_a{}^{\b'}K_{\b\b'}{}^\c h_\c{}^{\c'}
                - K_{a\b}{}^{\c'}\komma\hr
        \hfill\hbox{(ix)}&\hfill(\G^d)_{\a\b}m_d{}^c &=
                & (\G^c)_{\a\b}+(h_\a \G^c h_\b)\komma\hr
        \hfill\hbox{(x)}&\hfill 0 &=&  h_{(\a}(\G^{c'})_{\b)} \komma\hr
        \hfill\hbox{(xi)}&\hfill L_{(\a\b)}{}^\c
                +h_{(\b}{}^{\b'}K_{\a)\b'}{}^\c &=& 0\komma\hr
        \hfill\hbox{(xii)}&\hfill \woldelmod_{(\a}h_{\b)}{}^{\c'} &=
                & \ihalf({\G^c})_{\a\b}
                \chi_c{}^{\c'} + h_{(\a}{}^{\b'}K_{\b)\b'}{}^\c h_\c{}^{\c'}
                -K_{(\a\b)}{}^{\c'}\punkt\hr
}}
If we go through these equations we see that (i), (ii) and (iv)
contain no information for the fields but simply describe parts of
the torsion in the connection.
Equations (iii) and (vii) determine the remaining world-volume
torsion components in terms of the fields. Equation (v) does not generate
any new fields and thus
becomes an algebraic identity for the next-to-leading term in the
superfield ${m_a}^b$. Two, more manifest,
algebraic identities are (vi) and (xi). From (ix) and (x) we get
\eq{    h_\a{}^{\b'}=\fraction{6}(\G^{abc})_\a{}^{\b'}h_{abc}
}
and
\eq{
        m_a{}^b=\d_a{}^b-2k_a{}^b\komma
}
where $k_a{}^b=h_{acd}h^{bcd}$.
We note that putting
$\wT_{\a\b}{}^c=-i(\G^c)_{\a\b}$ implies that
\eq{    h_{[\a\b]}^{(ij)}=0\komma
}
which is identical to the on-shell constraint imposed in the NR formalism
of the previous sections.

In order to get the Dirac equation we take (xii):
\eq{
        \cec_{(\a\b)}{}^{\c'}=\Fraction{i}{2}(\G^c)_{\a\b}\chi_c{}^{\c'}
}
and trace the three free spinor indices in different ways to extract the
information.
By applying $(\G_d)^{\a\b}$ and
$(\G^d)_{\c'}{}^\b$ plus noting that
\eql{
        \cec_\b{}^{\c'}=\woldelmod h_\b{}^{\c'}
        -\half(\G^b{}_{c'})_\b{}^{\c'}\cec_b{}^{c'}
}{\krel}
we get
\eql{   \ki_a{}^{\c'}=-\Fraction{i}{4}(\G_a)^{\a\b}\cec_{\a\b}{}^{\c'}
}{\kifree}
and
\eql{   i\Bigl(\ki_c{}^{\c'}(\G^{cd})_{\c'\a}+\ki^d{}_\a\Bigr)=\fraction{2}
        (\G^b{}_{a'}\G^d)_\a{}^\b \cec_{\b b}{}^{a'}
        -\fraction{6}(\G^{abc}\G^d)_\a{}^\b\woldelmod_\b h_{abc}
}{\gammaone}
respectively. Now multiplying (\gammaone) by $(\G_d)_{\d'}{}^\a$, in order to
        get rid of $h_{abc}$, gives
\eql{   (\G^c)_{\d'\c'}\ki_c{}^{\c'}=\Fraction{i}{2}(\G^b{}_{a'})_{\d'}{}^\b
        \cec_{\b b}{}^{a'}\komma
}{\hremoved}
and if we use (\kifree) in (\hremoved) we get
\eq{    (\G^d{}_{a'})_{\d'}{}^\b \cec_{\b d}{}^{a'}=0\punkt
}
By comparing the two last equations we see that
\eq{    (\G^a)_{\a'\c'} \ki_a{}^{\c'}=0\komma
}
which is the Dirac equation.

In order to get the scalar and tensor equations of motion we take (viii):
\eql{
\woldelmod_\b\ki_a{}^{\c'}+Z_{a\b}{}^{\c'}=\cec_{a\b}{}^{\c'}\komma
}{\steq}
where
\eq{
Z_{a\b}{}^{\c'}:=L_{\b a}{}^d \chi_d{}^{\c'}+\chi_a{}^{\b'}
        K_{\b\b'}{}^{\c}h_\c{}^{\c'}\punkt
}
By using (\krel) in (\steq) we get
\eql{
\woldel_\b\ki_a{}^{\c'}+Z_{a\b}{}^{\c'}
=(\fraction{6}\G^{bcd}\woldelmod_ah_{bcd}-\half\G^b\G_{c'}\cec_{ab}{}^{c'}
)_\b{}^{\c'}\punkt
}{\eqmpre}
We now multiply (\eqmpre) by $(\G^{ae'})_{\c'}{}^\b$ and use the
Dirac equation,
which gives us the scalar equation
\eq{
\eta^{ab}\cec_{ab}{}^{c'}=\fraction{4}(\G^{ac'})_{\c'}{}^\b Z_{a\b}{}^{\c'}
	\punkt
}
If we instead multiply (\eqmpre) by $(\G^a\G_{ef})_{\c'}{}^\b$ (and again use
the Dirac equation) we get the tensor equation
\eq{
\woldelmod^ch_{abc}=\fraction{8}(\G^c\G_{ab})_{\c'}{}^\b Z_{c\b}{}^{\c'}\punkt
}
These equations of motion are analogous to the ones derived in 
ref. [\HSWFivebrane], and contain non-linearities of the same kind.

\section\Conclusions{Summary and conclusions}We have given a detailed 
account of the geometry involved in embeddings of supermanifolds into
supermanifolds. Special emphasis is put on the distinction between 
the different
geometric objects encountered, since confusing e.g. intrinsic and
induced geometry obscures the understanding of the formalism.
Two preferred parametrisations of the embedding matrix
in terms of matter fields, equations
(\Embeddingone) and (\Embeddingtwo) have been presented,
aiming towards distinct formulations of the world-volume field
theory, each one emphasising different properties of the theory.  

The second of these, referred to as the ``embedding formalism'',
investigated by Howe, Sezgin and West [\HSW], uses the torsion equation
(\teq) together with the geometric
``embedding condition'' $\te_\a{}^\ta=0$ in order to derive equations of
motion for the fields parametrising the embedding matrix (\Embeddingtwo).
The second formulation occurs in the theory of non-linear realisations
applied to the second supersymmetry (and the broken translations),
as advocated by Bagger and Galperin [\Bagger].
By using the second parametrisation (\Embeddingone) of the embedding matrix,
that formalism is rederived. 

We also described briefly the transformations involved in going from
one parametrisation to the other. Although it was straightforward to
show that these transformations exist, we did not examine them in detail.
As commented
on in section \Ibrane, the transformations,
eq. (\Minvariance), represent a kind of local symmetry inherent in the
definition of the embedding matrix, and it may be interesting to 
pursue the investigation further in order to extract information
from the field redefinitions. We remind that the transformations involve
not only the matter fields, but also the world-volume geometry.

We see two valuable aspects of this exercise. On one hand, the equivalence
of two seemingly different starting points is established, and it becomes
clear why they yield the same results (e.g. Born--Infeld dynamics).
On the second hand it casts some light on the embedding procedure in
explaining clearly why the obtained theory is one whose non-linearities
stem from the (non-linearly realised) symmetry under the target space
supersymmetry generators broken by the embedding.

The two parametrisations have been applied to a concrete case, namely 
the 5-brane of 7-dimensional supergravity. Here it was shown (in the
first of the parametrisations) that the embedding condition alone did
not provide enough information to put the theory on-shell.
An additional irreducibility constraint, completely analogous to the
one in the linear theory, had to be imposed. While this ``algebraic''
consideration became transparent in the language of non-linear realisations,
the torsion components here become so complicated that we find the extraction
of the field equations, though in principle possible,
quite non-transparent.
The second of the parametrisations, on the other hand, is quite suited
for finding the field equations (section \EqOM). In this case we did
not need to impose any additional constraint after the 
world-volume torsion was chosen to be that of conformal 6-dimensional 
supergravity. Since we could associate the irreducibility constraint
of our second formulation with the vanishing of a specific torsion
component, we conjecture that the choice of torsion in the second case
was more than a conventional one, so that the irreducibility constraint
is hidden in the vanishing of the $\g_{abc}$ part of the dimension-0
torsion component.

\acknowledgements The authors are grateful to Paul Howe
for commenting some of the calculations in ref. [\HSW]. The work of M.C. was
sponsored by the Swedish Natural Science Research Council.

\vfill\eject
\appendix{Notation and conventions}Since a number of different geometric
objects referring to different structures are encountered in this paper,
we try to summarise the notation in the following 
table\footnote*{Like the authors of ref. [{\xold}7]
we use the term ``intrinsic'' for the a priori defined
world-volume entities, but note that there is
an unfortunate disagreement on terminology. Mathematical literature may
use the term for what we call ``induced''.}.

\vskip3\parskip\noindent
\vbox{\tabskip=0pt\offinterlineskip\def\ttt{\noalign{\hrule}}
\halign{#&\vrule#\tabskip=4pt&#\hfil&\vrule width1pt#
        &\hbox to 1.8cm{\hfil#\hfil}&\vrule#
        &\hbox to 1.8cm{\hfil#\hfil}&\vrule#
        &\hbox to 1.8cm{\hfil#\hfil}&\vrule#
        &\hbox to 1.8cm{\hfil#\hfil}&\vrule#
        &\hbox to 1.8cm{\hfil#\hfil}&\vrule#
        \tabskip=0pt\cr\ttt
&height1pt&&&&&&&&&&&&\cr
&height11pt&&&\xrm world-volume&&\xrm world-volume&&&&&&Target&\cr
&height11pt&&&Intrinsic&&Induced
 &&Extrinsic&&Normal&&space&\cr
&height2pt&&&&&&&&&&&&\cr\noalign{\hrule height1pt}
&height2pt&&&&&&&&&&&&\cr
&height11pt&Metric&&$\wm$&&$\im$&&&&$\nm$&&$\tm$&\cr
&height2pt&&&&&&&&&&&&\cr\ttt
&height2pt&&&&&&&&&&&&\cr
&height11pt&Vielbein&&$\we_A$&&$\iE_A$&&&&$\nE_{A'}$&&$\tE_{\tA}$&\cr
&height2pt&&&&&&&&&&&&\cr\ttt
&height2pt&&&&&&&&&&&&\cr
&height11pt&Connection&&$\ww_A{}^B$&&$\iw_A{}^B$&&&&$\nw_{A'}{}^{B'}$
        &&$\tw_{\tA}{}^{\tB}$&\cr
&height2pt&&&&&&&&&&&&\cr\ttt
&height2pt&&&&&&&&&&&&\cr
&height11pt&Torsion&&$\wT^A$&&$\iT^A$&&$\eT_{AB}{}^{C'}$&&$\nT^{A'}$
        &&$\tT^{\tA}$&\cr
&height2pt&&&&&&&&&&&&\cr\ttt
&height2pt&&&&&&&&&&&&\cr
&height11pt&Curvature&&$\wr_A{}^B$&&$\ir_A{}^B$&&$\cec_{AB}{}^{C'}$
        &&$\nr_{A'}{}^{B'}$&&$\tr_{\tA}{}^{\tB}$&\cr
&height2pt&&&&&&&&&&&&\cr\ttt
&height2pt&&&&&&&&&&&&\cr
&height11pt&Exterior derivative&&$\wd$&&$\id$&&&&$\nd$&&$\td$&\cr
&height2pt&&&&&&&&&&&&\cr\ttt
&height2pt&&&&&&&&&&&&\cr
&height11pt&Canonical 1-form&&$\wco$&&$\ico$&&&&$\nco$&&$\tco$&\cr
&height2pt&&&&&&&&&&&&\cr\ttt
&height2pt&&&&&&&&&&&&\cr
&height11pt&Covariant derivative&&$\wdel$&&$\idel$&&&&$\ndel$&&$\tdel$&\cr
&height2pt&&&&&&&&&&&&\cr\ttt
}}\vskip2\parskip

\appendix{Spinors in 6 and 7 dimensions}The $D=7$ $\G$-matrices decompose as
\eq{(\G^a)_{\ol{\a \b}}=\mx{(\G^a)_{\a \b}&0\\
        0&(\bar{\G}^a)_{\a'\b'}}
}
and
\eq{(\G^{a'})_{\ol{\a \b}}=\mx{ 0&(\G^{a'})_{\a\b'}\\
                        (\G^{a'})_{\a'\b}&0\\}
}
with respect to the tangential and normal directions and they satisfy
\eq{    (\G^\ba)_{\ul{\a\b}}=(\G^\ba)_{\ul{\b\a}}\punkt
}
To raise and lower
composite indices we use
\eq{    C_{\ol{\a\b}}=C^{\ol{\a\b}}=\mx{0&1\\
                                    -1& 0}
                                =\mx{0& \d^{\a\b'}\\
                                     -\d^{\a'\b}&0}\komma
}
with the convention that
\eqa{   \psi^\baa=C^{\ol{\a\b}}\psi_\bbb\komma\\
        \psi_\bbb=\psi^\bbb C_{\ol{\b \a}}\punkt
}
The algebra is
\eq{    \{\G^\ba, \G^\bb\}=2\eta^{\ol{ab}}\komma
}
which implies that
\eqa{   \{\G^a,\bar{\G}^b\}:=&\G^a\bar{\G}^b+\G^b\bar{\G}^a
                =-2\eta^{ab}\d_\a{}^\b\komma\\
        \{\bar{\G}^a,\G^b\}:=&\bar{\G}^a\G^b+\bar{\G}^b\G^a
                =-2\eta^{ab}\d_{\a'}{}^{\b'}\komma\\
        \{\G^{a'},\G^{b'}\}:=&\G^{a'}\G^{b'}+\G^{b'}\G^{a'}=
                2\eta^{a'b'}\d_\baa{}^\bbb\komma\\
        \{\G^{a'},\G^b\}:=&\G^{a'}\G^b+\G^b\G^{a'}=0\komma\\
}

We split the 16 component indices according to
\eqa{   \psi_\a \rightarrow \psi_\a^i\komma\cr
        \psi_{\a'} \rightarrow \psi_i^\a\komma\cr
}
where after the split $\a$ is a Spin(1,5) index and $i$ is a SU(2) index.
For the $\G$-matrices this implies
\eq{    (\G^a)_\baa{}^\bbb=\mx{0&-(\G^a)_\a{}^{\b'}\\
                (\bar{\G}^a)_{\a'}{}^\b&0}
        \rightarrow
        \mx{0&-\e^{ij}(\c^a)_{\a\b}\\
                \e_{ij}(\bar{\c}^a)^{\a\b}&0}
}
and
\eq{    (\G^{a'})_\baa{}^\bbb= \mx{(\G^{a'})_\a{}^\b&0\\
                                0&-(\G^{a'})_{\a'}{}^{\b'}}
        \rightarrow              \mx{(\c^7)_\a{}^\b \d^i{}_j&0\\
                                0&-(\c^7)^\a{}_\b\d_i{}^j}\komma
}
where $\c^a$ are the 6-dimensional gamma matrices [\KT]. They satisfy
\eqa{   (\c^a)_{\a\b}=&-(\c^a)_{\b\a}\komma\\
        (\c_a)_{\a\b}(\c^b)^{\a\b}=&-4\d_a{}^b
}
and
\eq{    (\c^a)_{\a\b}(\c_a)_{\c\d}=-2\e_{\a\b\c\d}\punkt
}
Indices are raised
and lowered according to
\eqa{   \psi^i=&\e^{ij}\psi_j\komma\\
        \psi_i=&\psi^j\e_{ji}\\
}
and
\eqa{   \psi^{\a\b}=&\fraction{2}\e^{\a\b\c\d}\psi_{\c\d}\komma\\
        \psi_{\a\b}=&\fraction{2}\e_{\a\b\c\d}\psi^{\c\d}\punkt
}
Notice that we can only raise and lower Spin(1,5) indices in pairs.

\appendix{Some useful relations}In order to transform between
vector and spinor indices we need the following relations, following from
the lorentzian property of the $u$ matrices:
\eqa{   (\woldel u_\baa{}^\tcc)u_\tcc{}^\bbb&=-\fraction{4}(\G^\ba{}_\bb)_\baa
        {}^\bbb(\woldel u_\ba{}^\tc)u_\tc{}^\bb\komma\cr
        (\woldel u_\ba{}^\tc)u_\tc{}^\bb&=\Fraction{2}{$\SS \un$}
        (\G_\ba{}^\bb)_\bbb
        {}^\baa(\woldel u_\baa{}^\tcc)u_\tcc{}^\bbb\komma\cr
}
where $\un$ is the dimension of the target space 
spinor representation. If we take
into account
the split into tangential and normal indices we get
\eqa{   &(\woldel u_\a{}^\tcc)u_\tcc{}^\b
        =-\fraction{4}\Bigl((\G^a{}_b)_\a{}^\b(\woldel
        u_a{}^\tc)u_\tc{}^b
        +(\G^{a'}{}_{b'})_\a{}^\b(\woldel u_{a'}{}^\tc)u_\tc{}^{b'}\Bigr)
                \komma\cr
        &(\woldel u_\a{}^\tcc)u_\tcc{}^{\b'}
        =-\fraction{2}(\G^a{}_{b'})_\a{}^{\b'}
        (\woldel u_a{}^\tc)u_\tc{}^{b'}=-\fraction{2}(\G^{a'}{}_b)_\a{}^{\b'}
        (\woldel u_{a'}{}^\tc)u_\tc{}^b\komma\cr
        &(\woldel u_{\a'}{}^\tcc)u_\tcc{}^{\b'}
        =-\fraction{4}\Bigl((\G^a{}_b)_{\a'}{}^{\b'}(\woldel
        u_a{}^\tc)u_\tc{}^b
        +(\G^{a'}{}_{b'})_{\a'}{}^{\b'}(\woldel u_{a'}{}^\tc)u_\tc{}^{b'}
                \Bigr)\komma\cr
}
and
\eqa{   &(\woldel u_a{}^\tc)u_\tc{}^b
                =\Fraction{4}{$\SS \un$}(\G_a{}^b)_\b{}^\a(\woldel
        u_\a{}^\tcc)u_\tcc{}^\b
                =\Fraction{4}{$\SS \un$}(\G_a{}^b)_{\b'}{}^{\a'}(\woldel
        u_{\a'}{}^\tcc)u_\tcc{}^{\b'}\komma\cr
        &(\woldel u_{a'}{}^\tc)u_\tc{}^{b'}
                =\Fraction{4}{$\SS \un$}(\G_{a'}{}^{b'})_\b{}^\a(\woldel
        u_\a{}^\tcc)u_\tcc{}^\b
                =\Fraction{4}{$\SS \un$}(\G_{a'}{}^{b'})_{\b'}{}^{\a'}(\woldel
        u_{\a'}{}^\tcc)u_\tcc{}^{\b'}\komma\cr
        &(\woldel u_a{}^\tc)u_\tc{}^{b'}
                =\Fraction{4}{$\SS \un$}(\G_a{}^{b'})_\b{}^{\a'}(\woldel
        u_{\a'}{}^\tcc)u_\tcc{}^\b
                =\Fraction{4}{$\SS \un$}(\G_a{}^{b'})_{\b'}{}^\a(\woldel
        u_\a{}^\tcc)u_\tcc{}^{\b'}\punkt\cr
}

\vfill\eject
\refout

\end